\documentclass[letterpaper,11pt]{article}  
\usepackage{amsfonts}
\usepackage[pdftex]{graphicx}
\usepackage{amsmath,amssymb,amsthm}
\usepackage{natbib}
\usepackage{hyperref}
\usepackage{bm}
\usepackage{caption}
\usepackage{subcaption}
\usepackage{adjustbox,lipsum}
\usepackage{fancyhdr}
\usepackage{sectsty}
\usepackage{stmaryrd}
\usepackage{setspace}                      
\usepackage{booktabs}    
\usepackage{pdfsync}        
\usepackage[backgroundcolor=blue!40,linecolor=blue!40]{todonotes}
\usepackage{fancybox}
\usepackage{appendix}

\definecolor{cornellred}{RGB}{179,27,27} 
\definecolor{cornellblue}{RGB}{00,00,170}
\definecolor{cornellgrey}{RGB}{96,94,92}

\newtheoremstyle{myplain}
  {9pt}
  {9pt}
  {\itshape}
  {\parindent}
  {\scshape}
  {:}
  {.5em}
  {}
\newtheoremstyle{mydefinition}
  {9pt}
  {9pt}
  {\itshape}
  {\parindent}
  {\scshape}
  {:}
  {.5em}
  {}
\newtheoremstyle{myremark}
  {9pt}
  {9pt}
  {}
  {\parindent}
  {\scshape}
  {:}
  {.5em}
  {}
\theoremstyle{myplain}

\theoremstyle{mydefinition}

\theoremstyle{myremark}

\renewcommand{\cite}{\citet}

\usetikzlibrary{calc}
\def\centerarc[#1](#2)(#3:#4:#5){ \draw[#1] ($(#2)+({#5*cos(#3)},{#5*sin(#3)})$) arc (#3:#4:#5);}

\hypersetup{colorlinks=true, linkcolor=blue, citecolor=blue}                          
\numberwithin{equation}{section}

\begin{document}

\title{A Critical Assessment \\ of Some Recent Work on COVID-19\thanks{Thanks to Dominik Liebl, Chuck Manski, Francesca Molinari, and Christoph Rothe for conversations that helped shape this note.}}
\date{\today}
\author{J\"{o}rg Stoye\thanks{Department of Economics, 
Cornell University, stoye@cornell.edu.}}

\maketitle
\begin{abstract}
I tentatively re-analyze data from two well-publicized studies on COVID-19, namely the Charit\'{e} ``viral load in children" and the Bonn ``seroprevalence in Heinsberg/Gangelt" study, from information available in the preprints. The studies have the following in common:
\begin{itemize}
\item They received worldwide attention and arguably had policy impact.
\item The thrusts of their findings align with the respective lead authors' (different) public stances on appropriate response to COVID-19.
\item Tentatively, my reading of the Gangelt study neutralizes its thrust, and my reading of the Charit\'{e} study reverses it.
\end{itemize}
The exercise may aid in placing these studies in the literature. With all caveats that apply to $n=2$ quickfire analyses based off preprints, one also wonders whether it illustrates inadvertent effects of ``researcher degrees of freedom."
\end{abstract}

\vfill

\pagebreak
\onehalfspacing

\section{Introduction}
\label{sec:introduction}

This note takes a second look at two recent papers on COVID-19 released by leading German research groups. Both received widespread attention and arguably had policy impact. The first, by Christian Drosten's research group, was cited as evidence against opening schools; the second one, by Hendrick Streeck with coauthors, was cited as evidence in favor of a partial reopening of Germany. Both made important data collection efforts and then interpreted those data. I limit my analysis exclusively to this last step of interpretation, which is also the only aspect on which I can claim expertise. I find that, based on information available in the preprints, my own analysis would plausibly have had a neutral or even the opposite tendency in each case. I next elaborate this somewhat formally and then discuss implications.  

\section{The Charit\'{e} Study: Viral Load in Children}

\subsection{Background}
\begin{figure}[t]
\begin{adjustbox}{center}
\includegraphics[trim=4cm 8cm 4cm 8.8cm,clip=true]{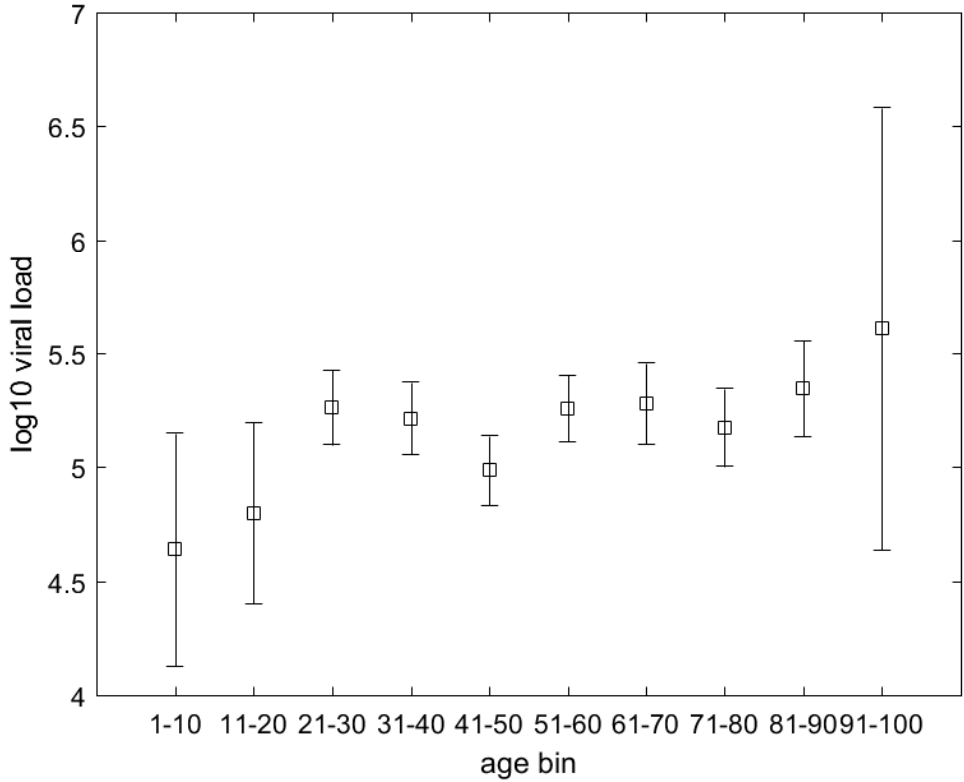}
\end{adjustbox}
\caption{Visualization of Table 2, panel 1, in the Charit\'{e} study. Error bars display $\pm 1.96$ standard errors. From the study's abstract: ``[T]hese data indicate that viral loads in the very young do not differ significantly from those of adults." \\ (Inspired by the image accompanying \cite{Textor20}.)}
\label{fig1}
\end{figure}

\cite[][henceforth ``the Charit\'{e} study"]{Drosten20} analyze viral load by age in a convenience sample of Covid-positive patients in Berlin. The aim of the investigation was to see whether epidemiological and anecdotal evidence of children being less contagious could be corroborated. The study received massive media coverage in Germany and abroad. Its data are visualized in Figure \ref{fig1}.\footnote{An obvious limitations is the use of a convenience sample with relatively few (presumably also relative to infected numbers) children. The authors are open about that, and I have nothing to say about it.} The seemingly strong association between age and viral load is corroborated by a nonparametric omnibus test of association. However, the paper next tests for significant differences across all $45$ pairwise combinations of age bins (and repeats the analysis with different binning). After adjustment for multiple hypothesis testing, no specific comparison is significant. This is reported in the abstract as follows: ``[T]hese data indicate that viral loads in the very young do not differ significantly from those of adults."

\subsection{Reanalysis}
Some disagreements with the analysis may come down to taste. The authors focus on a hypothesis test as deliverable of their analysis; I would have recommended a nonparametric mean regression with error bands, resulting in some estimated age effect. The main reason is that substantive significance does not equal statistical significance -- if we found a minuscule but highly significant effect, we would not care either.\footnote{Conversely, I presume that the apparent age effect discussed later, which corresponds to a factor of $3$, is clinically significant because the authors would otherwise not bother with tests.} Next, the core finding is that a certain hypothesis does \textit{not} get rejected. Thus, the analysis suggests absence of evidence, which is of course not evidence of absence. Although this distinction is central to how we teach hypothesis tests, including in medicine \citep{Altman485}, it is at most barely maintained in the paper (see the aforecited sentence) and completely lost in its perception. As a result, the data that are visualized in Figure \ref{fig1} fueled headlines like ``Covid patients found to have similar virus levels across ages" \citep{Bloomberg20} and ``Children are just as contagious as adults" \citep{Kieselbach20}.

More troublingly, I am not convinced that the evidence \textit{is} absent. To the statistically educated reader, the above headlines may suggest that the study tests, and fails to reject, $H_0$: ``Children have the same viral load as adults." It does not. It rather splits the sample into $10$ age bins and then simultaneously tests whether any of the $45$ possible pairwise combinations of age groups exhibit a significant difference in load. With proper adjustment for the possibility of false discoveries from $45$ tests, not one comparison shows up as significant. This is reported as main conclusion, even though it is at tension with a test reported elsewhere in the preprint that indicates \textit{some} difference across age bins (Kruskal-Wallis test, Table A3, $p=0.8\%$ for the bins in Figure \ref{fig1}). That is, the analysis finds a significant age effect -- it is merely that post-hoc comparison does not single out a specific one of $45$ pairs as culprit.

If one's take-home message is a finding of nonsignificance, one should make a good faith effort to apply a powerful test. That did not seem to happen here. On the contrary, artificially coarsening the age variable and, especially, pooling the comparison of interest with 44 comparisons that nobody inquired about is bound to sacrifice power. What would the result have been without these choices? To take an educated guess, I will now try to recover from the paper a test of $H_0$: ``Children have the same viral load as adults." To this purpose, I combine the first two and the remaining age bins of Figure \ref{fig1} to find means of $4.74$ and $5.21$ with $95\%$ confidence intervals of $[4.42,5.05]$ and $[5.15,5.27]$, respectively. A simple t-test rejects the null of equal means at a two-sided p-value of $0.4\%$.\footnote{Independent sample t-test using standard normal critical values. The numbers needed for these computations were backed out independently, with identical results, from each of the two panels of Table 2.} It would seem that the lower viral load in the youngest age groups is highly significant. What gives?

I reiterate that this simple test would not be my preferred analysis of the raw data. However, I am willing to defend it vis a vis the preprint's multiple nonparametric hypothesis tests. Specifically:
\begin{itemize}
\item The test only considers one rather than 45 hypotheses. I consider this a feature and not a bug. The data were collected specifically to investigate the relative viral load of children, and the methodological justification for pooling this comparison with 44 others is not elaborated. Indeed, one could arguably conduct a one-sided test and report a p-value of $0.2\%$.
\item The test is influenced by the authors' choice of bins and by my inspection of the visualization, so it strictly speaking involved an informal specification search. However, given the study's purpose, my own pre-data plan (conditionally on involving a simple hypothesis test to begin with) would have proposed more or less the same comparison. Hence, this should not a major concern. That said, taking the above p-value entirely at face value is not recommended, but it is also not necessary for my point.
\item The test is exact only if relevant true distributions are normal. This is clearly not the case, neither exactly (negative values are impossible) nor approximately (the paper reports corresponding tests). However, I compare means across two samples of size $127$ and $3585$, and the empirical distributions appear slightly skewed but well-behaved (see Figure 1 in the Charit\'{e} study). An appeal to the Central Limit Theorem would seem innocuous.

In fact, this consideration is more pressing for the Charit\'{e} study itself. They perform 45 simultaneous tests on relatively small bins and therefore encounter the one-two punch of (i) looking at the further-out-in-the-tails quantiles of sampling distributions that become relevant with Bonferroni-type adjustments and (ii) rather small sample cells. Indeed, $9$ of the tests involve the ``91-100" age bin, which contains $17$ observations. This makes it harder to appeal to normal approximations, yet two of the three multiple hypothesis tests reported in the paper do so anyway and essentially differ from my test only through adjusting for multiplicity of hypotheses.\footnote{I thank Dominik Liebl for alerting me to the parametric nature of their tests. \cite{Liebl20} also reports further inconsistencies in the analysis.}
\end{itemize}
This simple analysis is not meant to be conclusive. My only point is that the paper's non-significance result seems to be driven by researcher choices. For any conclusions beyond that, I would want to see the results of a nonparametric regression analysis that undoes the age binning of data. The closest approximation to this that I am aware of is the insightful renalysis by \cite{Held20}, which qualitatively agrees with mine. The ideal solution here would be publication of the data; while this might not be legally easy, I note that a lot could be done with literally only age and viral load.

\section{The Gangelt Study: Seroprevalence in Germany}

\subsection{Background}
\cite[][henceforth ``the Gangelt study"]{Streeck20} is an early seroprevalence study conducted in the small German town of Gangelt, which had witnessed a superspreader event. The study made headlines for estimating a local prevalence of $15\%$ and especially an IFR of $.4\%$, which was perceived as surprisingly low (and would certainly be considered on the very low end now). From the latter number, the authors also informally extrapolate to $1.8$ million infected Germans, an extrapolation that was correspondingly perceived as high, therefore arguably reassuring, but also implausible. The study received considerable attention that was apparently seeked out by the lead author, who vocally advocated for partial reopening of North Rhine-Westphalia.\footnote{The extrapolation is presented as somewhat of an afterthought in the preprint, which also has extremely valuable information on other issues, e.g., on within-household transmission. However, it is the extrapolation that made headlines and that was emphasized in the press release.}

\paragraph{Reanalysis.}
It has been widely remarked that, even if one feels confidence extrapolating Gangelt's IFR, one cannot extrapolate the CI to either nationwide IFR or nationwide prevalence. The reason is that Gangelt's fatality count was considered nonstochastic, but for the purpose of extrapolating results to Germany, it is a realization of $8$ ``successes" (scare quotes very much intended) in a large trial. For illustration, a simplistic $95\%$ confidence interval for Gangelt's \textit{expected} fatality count covers all integers from $4$ to $16$.\footnote{This CI just inverts the Poisson distribution. It is reported to illustrate that the issue is not negligible. See \cite{Rothe20} for a sophisticated analysis of the multilayered inference problem. Note also that I use the higher (=$8$) of two fatality counts offered in the preprint.}  
 
But the point estimator itself may not be the only plausible choice either. It implies that only $10\%$ of cases were detected nationwide, an implication that was picked up by the media \citep{Burger20} but also frequently called out as implausible. Indeed, the discovery rate in Gangelt was estimated to be $20\%$. Extrapolating from this number to Germany would imply an IFR of $.9\%$ rather than $.4\%$. What gives?

The discrepancy stems from the fact that, if we believe Gangelt to be representative of Germany, we have overdetermined but not quite consistent estimating equations. For sake of argument, say we have $8$ fatalities, $388$ officially confirmed cases, and $1956$ true infections in Gangelt, while we have $7928$ fatalities, $174975$ confirmed cases, and an unknown true prevalence in Germany.\footnote{The nationwide numbers were reported by Johns Hopkins University on 5/14. They implied a slightly lower case fatality rate at the time of the study.} By the rule of proportions, we could back out an estimate of nationwide prevalence either as $1956 \times 7928/8=1938396$ (an update of the study's preferred estimate) or as $ 1956\times 174975/388= 882090$ (extrapolating the undercount). The numbers differ by so much because the crude case fatality rate (CFR) equals $8/388=2.1\%$ for Gangelt and $7928/174975=4.5\%$ for Germany. Thus, assuming that the IFR (interpreted as population-level expectation) is actually the same, Gangelt either tested more or got really lucky. By extrapolating the local IFR, the study assumes the former but without discussing this assumption.

At the same time, proponents of extrapolating the undercount could point out that the underlying binomial samples of $8/388$ versus $7928/174975$ are of vastly different size. Should we really prioritize the former over the latter in extrapolation? Or was Gangelt indeed just lucky? This is marginally implausible in that the (exact binomial) two-sided p-value for the former sample being drawn from the latter urn is $2.6\%$. But modest demographic difference between Gangelt and Germany, or accounting for clustered sampling of both numerator and denominator (the sample units were households; the superspreading event apparently spared local nursing homes), could easily compensate for that. What's more, the fatality count meanwhile increased to $9$. While it is not clear that this fatality was among the infected during the study's time frame, adding it would change the aforementioned p-value to $5.2\%$. Also, it would mean that the local and nationwide numbers are consistent up to sampling error according to the more sophisticated analysis in \cite[][numerical claim verified in personal communication]{Rothe20}.

Again, my point is not that the study's implicit argument is necessarily false. Gangelt's undercount might indeed be atypically low: Being a hotspot, the town saw much more testing than other places. However, that does not logically imply that Gangelt had more testing \textit{relative to true prevalence}. My complaint is that this case was not made, and a different extrapolation that would align the study with other recent seroprevalence studies (and have shorter confidence intervals to boot) was not discussed. A balanced analysis might well provide both and also some form of interpolation. Partial identification would be one framework close to my heart, but the primary concern is transparency about how different identifying assumptions would interact with the data at hand. 

\section{Conclusion}

This is not a ``gotcha" paper. Both aforecited studies are carefully and competently executed, especially considering the extreme time pressure. They are transparent about computations, the computations appear correct, and I have no doubt that they were executed in good faith. Compared to some concurrent work by prominent U.S. researchers (e.g., as discussed by \cite{Gelman20}), both studies are models of good science. Also, there are obvious limitations to quickfire reanalysis without raw data, and everything I have to say is accordingly tentative. With that said, I stand by the following assessments: There are many good arguments against a quick reopening of schools, but the Charit\'{e} study does not add to them. Neither does the Gangelt study paint a substantially more optimistic picture on IFR than other recent seroprevalence studies.\footnote{These remarks weigh in on roughly opposing sides of the ``reopening" debate, so I hope it is clear that I have no axe to grind on that dimension. (Nor do I think that the public would or should be especially interested in my personal opinion on these matters.)}

With the due caution warranted by a sample size of $n=2$, it is of note that both studies were perceived as having a certain policy thrust; in both cases, this thrust correlated with public perception of some authors' stand on COVID-19 policy; and in neither case was it corroborated by my reading. That is my methodological point. Specific, reasonable but not objectively forced choices of statistical analysis --what is sometimes referred to as ``researcher degrees of freedom"-- informed results that align with researchers' public stands. I emphasize that I do not suggest any intent. However, even at the very top of the discipline, the effects of, for example, deciding when an analysis is complete cannot be assumed to just go away.

While I think that my comments on the specific studies are of interest, an important message to you --the reader-- and me is to redouble consideration for such effects in \textit{our} next study. On an immediately practical note, it illustrates that  a data repository for COVID-19-related research would be invaluable. Finally, it is a reminder of the importance of professional peer review. While both of the above studies were controversially discussed by experts right away, my impression is that critical voices gained any traction with the public only for the Gangelt one. In sum, this may be the rare case where rapid peer review of an academic paper would plausibly have affected the news cycle.
\newpage

\bibliography{covid}

\end{document}